\documentstyle[12pt]{article}
\begin{document}

\def\ve#1{\mid #1\rangle}
\def\vc#1{\langle #1\mid}

\newcommand{\p}[1]{(\ref{#1})}
\newcommand{\be}{\begin{equation}}
\newcommand{\ee}{\end{equation}}
\newcommand{\sect}[1]{\setcounter{equation}{0}\section{#1}}

\newcommand{\vs}[1]{\rule[- #1 mm]{0mm}{#1 mm}}
\newcommand{\hs}[1]{\hspace{#1mm}}
\newcommand{\mb}[1]{\hs{5}\mbox{#1}\hs{5}}
\newcommand{\Db}{{\overline D}}
\newcommand{\bea}{\begin{eqnarray}}
\newcommand{\eea}{\end{eqnarray}}
\newcommand{\wt}[1]{\widetilde{#1}}
\newcommand{\und}[1]{\underline{#1}}
\newcommand{\ov}[1]{\overline{#1}}
\newcommand{\sm}[2]{\frac{\mbox{\footnotesize #1}\vs{-2}}
	   {\vs{-2}\mbox{\footnotesize #2}}}
\newcommand{\prt}{\partial}
\newcommand{\eps}{\epsilon}

\newcommand{\R}{\mbox{\rule{0.2mm}{2.8mm}\hspace{-1.5mm} R}}
\newcommand{\Z}{Z\hspace{-2mm}Z}

\newcommand{\cd}{{\cal D}}
\newcommand{\cg}{{\cal G}}
\newcommand{\ck}{{\cal K}}
\newcommand{\cw}{{\cal W}}

\newcommand{\vj}{\vec{J}}
\newcommand{\vl}{\vec{\lambda}}
\newcommand{\vz}{\vec{\sigma}}
\newcommand{\vt}{\vec{\tau}}
\newcommand{\vw}{\vec{W}}
\newcommand{\poiss}{\stackrel{\otimes}{,}}

\def\l#1#2{\raisebox{.2ex}{$\displaystyle
  \mathop{#1}^{{\scriptstyle #2}\rightarrow}$}}
\def\r#1#2{\raisebox{.2ex}{$\displaystyle
 \mathop{#1}^{\leftarrow {\scriptstyle #2}}$}}

\renewcommand{\thefootnote}{\fnsymbol{footnote}}
\newpage
\vs{8}
\begin{center}
{\LARGE {\bf Graded Lie algebras, Representation theory,}}\\[0.6cm]
{\LARGE {\bf Integrable mappings and Systems}}\\[1cm]

\vs{8}

{\large A.N. Leznov }
{}~\\
\quad \\

{\em  {~$~^{(a)}$} Instituto de Investigaciones en Matem\'aticas
  Aplicadas y en Sistemas}~\\
{\em Universidad Nacional Aut\`onoma de M\`exico} \\
{\em Apartado Postal 48-3, 62251 Cuernavaca,    \\
     Morelos, M\'exico} \\

\vspace{2mm}

{\em {~$~^{(b)}$ Institute for High Energy Physics,}}\\
{\em 142284 Protvino, Moscow Region, Russia}\\

\vspace{2mm}

{\em {~$~^{(c)}$} Bogoliubov Laboratory of Theoretical Physics,
JINR,}\\
{\em 141980 Dubna, Moscow Region, Russia}\\

\end{center}
\vs{8}

\begin{abstract}

A new class of integrable mappings and chains is introduced.
Corresponding $(1+2)$ integrable systems invariant, with respect
to such discrete transformations, are presented in an explicit form.
Their soliton-type solutions are constructed in terms of
matrix elements of fundamental representations of semisimple
$A_n$ algebras for a given group element. The possibility of
generalizing this construction to multi--dimensional case is
discussed.

\end{abstract}

\vfill

{\em E-Mail:\
leznov@ce.ifisicam.unam.mx }
\newpage
\pagestyle{plain}
\renewcommand{\thefootnote}{\arabic{footnote}}
\setcounter{footnote}{0}

\section{Introduction}

In an old paper \cite{l1} of the author the effective method
(based on two-dimensional zero-curvature condition) was proposed
for constructing of exactly integrable systems in two dimensions
together with their general solutions. During the last fifteen years
situation became more clear and understandable.

It turned out that by the method of \cite{l1} it is possible to
construct the
integrable mappings \cite{FAL},\cite{l} responsible for the existence
of the
hierarchies of integrable systems. Each equation of a given hierarchy
is
invariant with respect to the transformation of the corresponding
mapping
(or substitution).

Moreover, it has became clear that the formalism of $L-A$ pair is not
the principal point of the whole construction.
There exists a more direct way for obtaining the integrable mappings
and
technique of the $L-A$ pair is not more than one of its
consequences.
The situation with respect to the explicit solution of the quantum
two-dimensional Toda lattice \cite{FL} for Heisenberg
operators is the most important argument for the necessity of the
development
of a new way disconnected with the representation of the zero
curvature.
In letter case the $L-A$ representation is absent at all and
nevertheless
the method of the present paper works as well. In this connection it
is
necessary remind also about the general solution of periodical Toda
lattice,
which can be obtained in the form of absolutely  convergent infinite
series
without any using of the $L--A$ pair formalism, but with the help of
below construction \cite{LSM}.

The scheme proposed in the present paper is the following one.

At the first step we introduce two integrable in quadratures
equations of
$S$--matrix type for the two group-valued functions, depending on
two different arguments. The coefficient functions of these equations
are determined by the structure of the corresponding Lie algebra and
the choice of definite grading in it. We make no difference between
algebra and super-algebra cases, recalling  only that even (odd)
elements
of super-algebras are always multiplied by even (odd) elements of
the Grassman space.

By help of these two group elements we construct a new composite one.
Arising relations of equivalence between its matrixes elements lead
to
integrable substitution.

At the second step we assume additional dependence on arbitrary
functions,
determining general solution of the integrable substitution (with the
fixed ends), on some "time--like" parameters.
This is achieved by two additional equations for the above mentioned
group
elements $M^{\pm}$ in such a way that the condition of their
selfconsistency leads to finding the explicit dependence of arbitrary
functions on both the space and time like parameters.
With the help of these time parameters we present the hierarchy of
completely integrable systems each one of which is invariant with
respect
to the transformation of the integrable mapping (constructed at the
first
step).

At last, (this is the third step which is, in fact, not in the close
connection with the previous ones) we observe that in the framework
of the method  \cite{l1} there exists some hidden previously omitted
non--trivial possibility for the generalization of the whole
construction
on the multi-dimensional case
\footnote{About the other possibility to enlarge the class of
integrable
systems which was not discussed in \cite{l1}, see \cite{dls}.}.

What is the most remarkable in this approach is the fact
that dimension of the possible "multi-generalization" is uniquely
determined by the properties of the algebra and the choice of the
grading in it. Sometimes "multi-generalization" is equivalent
to the trivial change of variables, sometimes it leads to
nontrivial new possibilities. Nevertheless, in all of the cases
it is possible to obtain only particular (but not the general)
solutions of arising in this way systems and equations.

The present paper is organized in the following way. In section 2 we
briefly repeat the content of first section of the paper \cite{l1}
but in ``opposite direction'' compared to the original.
In section 3 for convenience of the reader, we present the most
important for further consideration results from the theory of
representation of (super) semisimple algebras and  groups.
In section 4 we show how to avoid the relatively cumbersome procedure
of resolution of the Gauss decomposition (in the cases when it is
equivalent to solution of the whole problem); we
present the integrable systems together with their general solutions
in terms
of the matrix elements of the various fundamental representations of
semisimple algebras. In this way we construct  $UToda(m_1,m_2)$
integrable mappings or substitutions. In section 5 we demonstrate the
way of
introducing of the evolution parameters in order to obtain the
hierarchies of
integrable systems invariant with respect to transformation of
the constructed integrable mappings. In section 6 we discuss the
possibility
containing in this construction,  for its generalization to
the multi-dimensional case (with not arbitrary dimensions!).
The concluding remarks are concentrated in section 7.

\section{Moving in the opposite direction}

Let us have some arbitrary finite dimensional graded algebra $\cal
G$.
This means that $\cal G$ may be represented as a direct sum of
subspaces with
the different grading indexes
\begin{equation}
  {\cal G}=\left(\oplus^{N_-}_{k=1} {\cal G}_{-\frac{k}{2}}\right)
  {\cal G}_0 \left(\oplus^{N_+}_{k=1}{\cal G}_{\frac{k}{2}}\right).
\label{GR}
\end{equation}
The generators with the integer graded indexes are called bosonic,
while
with half-integer indexes -- the fermionic ones. Positive (negative)
grading
corresponds to upper (lower) triangular matrices.

Let $M_+(y),M_-(x)$ be the elements of the group (when it exists)
corresponding to algebra (\ref{GR}) and only some solutions of the
equations of $S$-matrix type in some given finite dimensional
representation
of initial algebra $\cal G$:
\begin{equation}
\frac {\partial M_-}{\partial x}=L^{m_1}_-(x) M_-=\sum_{s=0}^{m_1}
A^{-s}(x)
M_-,
\quad \frac {\partial M_+}{\partial y}=L^{m_2}_+(y)
M_+=\sum_{s=0}^{m_2}
B^{+s}(y) M_+ \label{ME}
\end{equation}
where $A^{-s},B^{+s}$ are arbitrary functions of their arguments
taking
values in corresponding graded subspaces; $s$ is an integer or
half-integer number.

Let us introduce the group element $K$
\begin{equation}
K=M_+M_-^{-1}. \label{E}
\end{equation}
By the logic of Ref.\ \cite{l1} it is necessary to represent $K$ in
the form of the Gauss decomposition
$$
K=M_+M_-^{-1}=N_-^{-1} g_0 N_+
$$
($N_{\pm}$ are elements of positive (negative) nilpotent subgroups,
$g_0$ -- of the group with the algebra of the zero subspace)
and to consider the group  element $G$
\begin{equation}
G=N_-M_+=g_0 N_+M_- \label{EE}
\end{equation}
As a direct consequence of above definitions and equations for
$M^{\pm}$
elements (\ref{ME}) we obtain the following relations:
$$
G_xG^{-1}=(N_-)_x N_-^{-1}=\sum_s^{m_2} R^{(-s}(x,y)
$$
\begin{equation}
{} s={1\over 2}, 1, {3\over 2}, 2,......\label{LA}
\end{equation}
$$
G_yG^{-1}=(g_0)_y g_0^{-1}+g_0 (N_+)_yN_+^{-1} (g_0)^{-1}=(g_0)_y
g_0^{-1}+
\sum_s^{m_1} R^{(+s}(x,y)
$$
The Maurer--Cartan identity applied to (\ref{LA}) leads to the
equations of
exactly integrable system with the general solution determined
by (\ref{EE}) and the above formulae.

All that has been said so far was the almost literal repetition of
\cite{l1},
but in the opposite direction to the original one.
The main technique difficulty under such a approach
consists in explicit resolving of the Gauss decomposition -- finding
from
(\ref{EE}) group elements $N_{\pm},g_0$ when $M_{\pm}$ are known.
This is
sufficiently cumbersome problem under direct attempts of its
solution.

At this place we would like to emphasize, that connection between the
equations (\ref{ME}) and the further chain of equalities connected
with the
element $K$ leading to zero-curvature
representation encoded in (\ref{LA}) is true only in the case when
the algebra
$\cal G$ can be integrated up to the corresponding group. In the case
of Lie
algebras as it well-known it is always possible \cite{E}. But there
are exist
many other interesting problems ( and among them exactly the cases of
the
quantum Toda chain and general solution of the periodically one
mentioned in
introduction \cite{FL}, \cite{LSM}), when solution of equations
(\ref{ME})
may be presented in quadrature but to pass to representation of
zero-curvature
is impossible due to the absent of the corresponding group element.

Fortunately, there exists a more direct way allowing us,
using the definition of the element $K$ (not always having the group
ones
sense), to reconstruct the form of equations of the integrable system
as well as its general solution. We will demonstrate this way
applying it
to the case of semisimple algebras  when it is possible to present in
explicit
form as integrable mappings and as their general solutions ( in the 
interrupted version) in terms of matrix elements of the various
fundamental
representations.

Exactly the same technique is applicable to quantum version of two-
dimensional Toda lattice and general solution of the periodical one, 
when corresponding group element is absent
( simultaneously with the rigorous representation of
zero--curvature).

\section{Some facts from the representation theory
of the semi\-simple algebras}

In this section we restrict ourselves to the case of semisimple
algebras.
In the general case the grading operator $H$ may be presented as
linear
combination of elements of commutative Cartan subalgebra (taking  the
unit or zero values on the generators of the simple roots in the
form):
\begin{equation}
H={\sum}^r_{i=1} (K^{-1}c)_i h_{i}
\label{cartan1}
\end{equation}
Here, $(K^{-1})_{j,i}$ is the inverse Cartan matrix,
$K^{-1}K=KK^{-1}=I$
and $c$ the column consisting of zeros and unities in arbitrary
order.

As usually the generators of the simple roots $X^{\pm}_i$
(raising,lowering
operators) and Cartan elements $h_i$  satisfy the system of
commutation
relations:
\begin{eqnarray}
[h_i , h_j]=0, \quad [h_i,X^{\pm}_j]=\pm K_{ij}X^{\pm}_j, \quad
[X^{+}_i,X^{-}_j\}={\delta}_{i,j} h_j, \quad (1 \leq i,j \leq r),
\label{aa6}
\end{eqnarray}
where $K_{ij}$ is the Cartan matrix, and the brackets $[,\}$ denote
the graded commutator, r- is the rank of the algebra.

The highest vector $\ve{j}$ ($\vc{j} \equiv \ve{j}^{\dagger}$) of
the $j$--th fundamental representation possesses the following
properties:
\begin{eqnarray}
X^{+}_i\ve{j}=0, \quad h_i\ve{j}={\delta}_{i,j}\ve{j}, \quad
\vc{j}\ve{j}=1.
\label{high}
\end{eqnarray}
The representation is exhibited by repeated applications of the
lowering operators $X^{-}_i$ to the $\ve{j}$ and
extracting all linear-independent vectors with non-zero norm. Its
first
few basis vectors are
\begin{eqnarray}
&& \ve{j}, \quad X^{-}_j\ve{j},  \quad X^{-}_i X^{-}_j\ve{j},\quad
K_{i,j}\neq 0,\quad i\neq j
\label{vectors}
\end{eqnarray}

In the fundamental representations, matrix elements of the arbitrary
group element $G$ satisfy the following important identity
\footnote{Let us remind the definition of the
superdeterminant, $sdet \left(\begin{array}{cc} A, & B \\ C, & D
\end{array}\right) \equiv det (A-BD^{-1}C ) (det D)^{-1}$.}
\cite{ls0}
\begin{eqnarray}
sdet \left(\begin{array}{cc} \vc{j}X_j^+GX_j^-\ve{j}, &
\vc{j}X_j^+G\ve{j} \\ \vc{j}GX_j^-\ve{j}, & \vc{j} G \ve{j}
\end{array}\right) = {\prod}^r_{i=1,i\neq_j}\vc{i} G
\ve{i}^{-K_{ji}},
\label{recrel}
\end{eqnarray}
where $K_{ji}$ are the elements of the Cartan matrix. The identity
(\ref
{recrel}) represents the generalization of the famous Jacobi identity
connecting determinants of (n-1), n and (n+1) orders of some special
matrixes to the case of arbitrary semisimple Lee super-group. As we
will see
in the next section, this identity is so important at the deriving of
the
integrable mappings, that one can even say that it is responsible for
their
existence. We conserve for (\ref{recrel}) the name of the first
Jacobi
identity. Besides (\ref{recrel}), there exists no more less important
independent identity \cite{l}:
$$
K_{i,j} (-1)^P{\vc{j}X_j^+X_i^+ G \ve{j}\over \vc{j} G
\ve{j}}+K_{j,i}
{\vc{i}X_i^+X_j^+ G \ve{i}\over \vc{i} G \ve{i}}+
$$
\begin{equation}
K_{i,j}K_{j,i} (-1)^{jP}{\vc{j}X_j^+ G \ve{j}\over \vc{j} G \ve{j}}
{\vc{i}X_{i}^+G \ve{i}\over \vc{i} G \ve{i}}=0,\quad K_{i,j}\neq 0
\label{J2}
\end{equation}
which will be called as a second Jacobi identity. This identity is
responsible (in the above sense) for the fact of existence of
hierarchy
of integrable systems each one of which is invariant with respect to
transformations of constructed integrable mapping.

As from (\ref{recrel}) either from (\ref{J2}) it is possible to
construct
many useful recurrent relations which will be used under further
consideration.

\section{Integrable mappings and chains}

In this section using the apparat of the previous one we will show
that
matrix elements of group element $K$ satisfy the closed system of
equations
of equivalence which can be interpreted as the integrable mapping or
the
chain like system with the known general solution. We restrict
ourselves
by the case $A_n$ semisimple algebra with the main embedding in it.
All other
cases try for their consideration a little more cumbersome ( only
from the
technical point of view) calculations
\footnote{The simplest example of such calculations for the case of
arbitrary
semisimple algebra reader can find in the Appendix of this paper.}.

Under the main embedding the grading operator (\ref{cartan1}) takes
unity
values on all generators of the simple roots. Generators of $\pm
m$-graded
subspaces have the form
$$
Y_i^{\pm m}=[X^{\pm}_{i+m-1}[...[X^{\pm}_{i+1},X^{\pm}_i]..]]
$$

The action of right "Lagrangian"  $\tilde L^{m_1}_-$ (\ref{ME}) (sign
$\tilde{}$ means that terms with the graded index zero are
interrupted from
$L^{m_1}_-$) on the state vector $\ve{i}$ may be presented in form
\begin{equation}
\tilde L^{m_1}_- \ve{i}=\sum_{n=1}^{m_1} \sum_{s=0}^{n-1}(-1)^s
\phi_{i-s}^n
T^-_s(X^-_{i-1}) T^+_{n-s-1}(X^-_{i+1}) X^-_i \ve{i}\label{AL}
\end{equation}
where
$$
T^+_m(X^-_i)= T^+_{m-1}(X^-_{i+1}) X^-_i\quad T^+_0(X^-_i)=1,\quad
T^-_m(X^-_i)= T^-_{m-1}(X^-_{i-1}) X^-_i\quad T^-_0(X^-_i)=1
$$
To understand (\ref{AL}) is not so difficult. In fact, in  the
$m$-graded
subspace there are exactly $m-1$ generators containing generator of
the given
simple root $i$. In connection with (\ref{vectors}) the different
from zero
contribution may be arised only in the case if this generator occurs
on the 
last place. Exactly about this tell us formula (\ref{AL}).

By the same reasons the action of "Lagrangian" $\tilde L^{m_2}_+$ on
the left
state vector $\vc{i}$ takes the form
\begin{equation}
\vc{i} \tilde L^{m_2}_+ =\vc{i} X^+_i \sum_{n=1}^{m_2}
\sum_{s=0}^{n-1}(-1)^s
\bar \phi_{s-i}^n R^-_s(X^-+{i-1}) R^+_{n-s-1}(X^+_{i+1})
\label{AL1}
\end{equation}
where
$$
R^+_m(X^+_i)=X^+_i R^+_{m-1}(X^+_{i+1})\quad R^+_0(X^+_i)=1,\quad
R^-_m(X^+_i)=X^+_i R^-_{m-1}(X^+_{i-1})\quad R^-_0(X^+_i)=1
$$
The following definitions will be used:
\begin{equation}
\bar \alpha_i^{\pm m}={\vc{i} R^{\pm}_m (X^+_i) K \ve{i}\over \vc{i}
K
\ve{i}},\quad \alpha_i^{\pm m}={\vc{i} K T^{\pm}_m(X^-_i)\ve{i}\over
\vc{i} K
\ve{i}}\label{A}
\end{equation}
$$
< i >\equiv \vc{i} K \ve{i},\quad \theta_i\equiv
{< i-1 > < i+1 >\over < i >^2}
$$

In what follows for concrete calculations we will use many times the
same
trick. The matrix element $\vc{a} X^+_l G X^-_k \ve{b}$, were
$\vc{a},\ve{b}$
are arbitrary state vectors of some representation of the group may
be
rewritten as:
\begin{equation}
\vc{a} X^+_l G X^-_k \ve{b}\equiv (\hat X^+_l)_{left} (\hat X^-_k)_
{right} \vc{a} G  \ve{b}\label{T}
\end{equation}
where now $(\hat {X^+_l})_{left}, (\hat {X^-_k})_{right}$ are the
corresponding generators of the left (right) regular representation
acting
on group element $G$. This approach allows to lead the most part of
the
calculations below only to repeatedly using of the first and
second Jacobi identities (\ref{recrel}),(\ref{J2}).

As a first example let us consider the proof of the following
recurrent
relations for function $Q^{\pm}_{a,b;i\pm 1}\equiv (R^{\pm}_a(X^+_
{i\pm 1})_l \vc{i\pm 1} K
\ve{i\pm 1} (T^{\pm}_b(X^-_{i\pm 1}))_r$ :
$$
Q^{\pm}_{a,b;i\pm 1}=(R^{\pm}_{a-1}(X^+_ {i\pm 2})_r
\vc{i\pm 1} X^+_{i\pm 1} K X^-_{i\pm 1} \ve{i\pm 1}
(T^{\pm}_{b-1}(X^-_{i\pm 2}))_r=
$$
\begin{equation}
{\vc{i} K \ve{i}\over \vc{i\pm1} K \ve{i\pm 1}}Q^{\pm}_
{a-1,b-1;i\pm 2}+\vc{i\pm1} K \ve{i\pm 1} \bar \alpha^{\pm a}_{i\pm
1}
\alpha^{\pm b}_{i\pm 1}\label{RR}
\end{equation}
In the process of performing the last transformations we have
expressed the
matrix element $\vc{i\pm 1} X^+_{i\pm 1} K X^-_{i\pm 1} \ve{i\pm 1} $
with
the help of the first Jacobi identity (\ref{recrel}) and used a
definition for
the $ \alpha,\bar \alpha$ functions (\ref{A}).

Now we are ready to pass to calculation of the equivalence relations
among
introduced above values.

We present here the main typical steps of calculation the derivative
of the
$\bar \alpha$ (\ref{A}) functions with respect to $x$ coordinate. We
have
consequently:
$$
(\bar \alpha^{+m}_j)_x=(\vc{j} K \ve{j})^{-2}
det \left(\begin{array}{cc}
\vc{j}R^+_m(X_j^+) K \tilde L^{m_1}_-\ve{j}, & \vc{j}R^+_m(X_j^+) K
\ve{j} \\
		  \vc{j} K \tilde L^{m_1}_-\ve{j}, & \vc{j} K \ve{j}
\end{array}\right) =
$$
$$
(\vc{j} K \ve{j})^{-2}(\hat R^+_{m-1}(X_{j+1}^+)_l \sum_{n=1}^{m_1}
\sum_{s=0}^{n-1}(-1)^s \phi_{s-i}^n (\hat T^-_s(X^-_{i-1}))_r
(\hat T^+_{n-s-1}(X^-_{i+1}))_rú
$$
$$
(\vc{j-1} K \ve{j-1}\vc{j+1} K \ve{j+1})
$$
Up to now in the process of the last evaluation we have used many
times the
mentioned above trick (\ref{T}) and first Jacobi identity
(\ref{recrel}).
Further evaluation is connected with the repeatedly application of
the 
recurrent relation (\ref{RR}) to the expression
$$
(\hat R^+_{m-1}(X_{j+1}^+)_l(\hat T^+_{n-s-1}(X^-_{i+1}))_r(\vc{j+1}
K
\ve{j+1})
$$
and following from definition (\ref{A}) relation:
$$
(\hat T^-_s(X^-_{i-1}))_r(\vc{j-1} K \ve{j-1})=(\vc{j-1} K
\ve{j-1})\alpha^
{-s}_{j-1}
$$
Performing all calculations having pure algebraic character we come
to
final expression for interesting for us derivative:
\begin{equation}
(\bar \alpha^{+m}_j)_x=\sum_{q=0}^{m-1}\Theta^{+q}_i p^{(q+1}_i \bar
\alpha^
{m-1-q}_{i+q+1} \label{FFE}
\end{equation}
where
$$
p^{(r}_i=\sum_{n=1}^{m_1} \sum_{s=0}^{n-1}(-1)^s \phi_{i-s}^n
\alpha^{-s}_
{i-1} \alpha^{n-s-r}_{i+r}\quad \Theta^{\pm p}=\prod_{r=0}^p
\theta_{i\pm r},
\quad p^{(m_1}_i \equiv 1,\quad p^{(r}_i=0,\quad m_1+1\leq r
$$
By the same technique we obtain also
$$
(\bar \alpha^{-m}_i)_x=\sum_{q=0}^{m-1}(-1)^q \Theta^{-q}_i
p^{(q+1}_{i-q}
\bar \alpha^{-(m-1-q)}_{i-q-1}
$$
and corresponding expressions for derivatives of $\alpha$--functions
with
respect to variable $y$ :
\begin{equation}
(\alpha^{+m}_j)_y=\sum_{q=0}^{m-1}\Theta^{+q}_i \bar p^{(q+1}_i
\alpha^
{m-1-q}_{i+q+1} \quad
(\alpha^{-m}_i)_y=\sum_{q=0}^{m-1}(-1)^q \Theta^{-q}_i \bar
p^{(q+1}_{i-q}
\bar \alpha^{-(m-1-q)}_{i-q-1} \label{FEE}
\end{equation}
where
$$
\bar p^{(r}_i=\sum_{n=1}^{m_2} \sum_{s=0}^{n-1}(-1)^s \bar
\phi_{s-i}^n
\bar \alpha^{-s}_{i-1} \bar \alpha^{n-s-r}_{i+r},\quad
\quad \bar p^{(m_2}_i \equiv 1,\quad \bar p^{(r}_i=0,\quad m_2+1\leq
r
$$
Now we describe the main steps of calculations of the mixed
derivative
$\frac {\partial^2 \ln (\vc{i} K \ve{i})}{\partial x\partial y}$.
Using
the main equations (\ref{ME}), it is possible (similar to
calculations above)
to present this derivative in the form of determinant of the second
order
and apply to it the first Jacobi identity (\ref{recrel}). In such way
we
obtain:
$$
\frac {\partial^2 \ln (\vc{i} K \ve{i})}{\partial x\partial y}=< i
>^{-2}
\sum_{n=1}^{m_1}\sum_{s=0}^{n-1}
\sum_{m=1}^{m_2}\sum_{t=0}^{m-1}(-1)^{s+t}
\phi^{(n}_{i-s} \bar \phi^{(m}_{i-t}ú
$$
$$
(\hat R^+_{m-t-1}(X^+_{i+1}))_l
(\hat T^+_{n-s-1}(X^-_{i+1}))_r \vc{j+1} K \ve{j+1}ú
$$
$$
(\hat R^-_t(X^+_{i-1}))_l
(\hat T^-_s(X^-_{i-1}))_r \vc{j-1} K \ve{j-1} =
$$
Application to both factors of the last sum recurrent relations
(\ref{RR}) allows to transform it to the form:
$$
\sum_{n=1}^{m_1}\sum_{s=0}^{n-1}
\sum_{m=1}^{m_2}\sum_{t=0}^{m-1}(-1)^{s+t}
\phi^{(n}_{i-s} \bar \phi^{(m}_{i-t} (\theta_i)^{-1} \sum_{p=0}
\Theta^{-p}
_{i-1} \bar \alpha^{-(t-p)}\alpha^{-(s-p)} \sum_{q=0}
\Theta^{+q}_{i+1}
\bar \alpha^{(m-t-q-1)}\alpha^{(n-s-q-1)}=
$$
and at last changing in the last sum the order of summation we come
to the
final expression:
$$
(\theta_i)^{-1} \sum_{p=0,q=0}^{p+q\leq Min\{ m_1-1,m_2-1 \}}
\Theta^{-p}_i \Theta^{+q} p^{(p+q+1}_{i-p} \bar p^{(p+q+1}_{i-p}
$$
where all functions involved are defined above.

The knowledge of the explicit expressions for derivatives of $\alpha$
and
$\bar \alpha$ functions with respect correspondingly to $ y,x$
coordinates
allow to calculate the derivatives of $p^{(r}_i$ and $\bar p^{(r}_i$
functions and obtain the closed system of equations ( or identities)
for
unknown functions $p^{(r}_i, \bar p^{(r}_i$ and $< i >$ (with the 
corresponding boundary conditions):
$$
\frac {\partial \bar p^{(r}_i}{\partial x}=\sum_{q=1}^{m_2-r}
(\Theta^{+(q-1)
}_{i+r} p^{(q}_{i+r}\bar p^{(q+r}_i-\Theta^{-(q-1)}_{i-1}
p^{(q}_{i-q}\bar p^
{(q+r}_{i-q})
$$
\begin{equation}
\frac {\partial^2 \ln (\vc{i} K \ve{i})}{\partial x\partial
y}=\theta_i^{-1}
\sum_{p=0,q=0}^{p+q\leq Min\{ m_1-1,m_2-1 \}} \Theta^{-p}_i
\Theta^{+q}_i
p^{(p+q+1}_{i-p}\bar p^{(p+q+1}_{i-p}
\label{SUP}
\end{equation}
$$
\frac {\partial p^{(r}_i}{\partial y}=\sum_{q=1}^{m_1-r}
(\Theta^{+(q-1)}_
{i+r} \bar p^{(q}_{i+r} p^{(q+r}_i-\Theta^{-(q-1)}_{i-1}\bar
p^{(q}_{i-q}
p^{(q+r}_{i-q})
$$

In paper \cite{l} the system (\ref{SUP}) was called as
UToda$(m_1,m_2)$ system and we preserve for it this name, keeping in
mind necessary modifications which was done in \cite{dls}.

We can forget about the boundary conditions and consider the lattice
system (\ref{SUP}) as infinite one, where index $i$ takes all natural
values (positive and negative ones).

In this case (\ref{SUP}) can be considered as some mapping -- the law
with the help of which some number of initial functions are connected
with
the same number of the finally ones. We would like to clarify
situation
on the concrete examples of UToda$(m_1,m_2)$ lattices with the lowest
numbers of $m_1,m_2$.

\subsection{UToda$(1,1)$}

In this case (\ref{SUP}) is equivalent to the chain of equations for
the
single unknown function $< i >$ or $\theta_i$ ( there are many other
equivalent forms of the usual Toda lattice):
$$
\frac {\partial^2 \ln < i >}{\partial x\partial y}={< i-1 >< i+1
>\over
< i >^2},\quad \frac {\partial^2 \ln \theta_i}{\partial x\partial y}=
\theta_{i+1}-2\theta_i+\theta_{i-1}
$$
The initial functions in this case are  the two functions
$\phi_1=\theta_i,
\phi_2=\theta_{i-1}$. The final ones are $\tilde \phi_1=\theta_{i+1},
\tilde \phi_2=\theta_i$ and the corresponding mapping takes the form:
$$
\tilde \phi_1=\frac {\partial^2 \ln \phi_1}{\partial x\partial
y}+2\phi_1-
\phi_2,\quad \tilde \phi_2=\phi_1
$$

\subsection{UToda$(1,2)$}

In this case (\ref{SUP}) takes the form of two chain equation for two
unknown
functions $\theta_i,p^{(1}_i$ in each point of the lattice:
$$
\frac {\partial^2 \ln \theta_i }{\partial x\partial y}=
\theta_{i+1}p^{(1}_{i+1}-2\theta_i p^{(1}_i+\theta_{i-1}
p^{(1}_{i-1},\quad
\frac {\partial p^{(1}_i}{\partial y}=\theta_{i+1}-\theta_{i-1}
$$
The last system may be rewritten in the mapping form. Four initial
functions
$\phi_1=\theta_i, \phi_2=\theta_{i-1}, \phi_3=p^{(1}_i,
\phi_4=p^{(1}_{i-1}$
are connected with the four final ones
$\tilde \phi_1=\theta_{i+1}, \tilde \phi_2=\theta_i, \tilde
\phi_3=p^{(1}_
{i+1}, \tilde \phi_4=p^{(1}_i$ with the help of the rule:
$$
\tilde \phi_4=\phi_3,\quad \tilde \phi_3 \tilde \phi_1=\frac
{\partial^2
\ln \phi_1}{\partial x\partial y}-\phi_2 \phi_4+2\phi_1 \phi_3,\quad
\tilde \phi_2=\phi_1,\quad \tilde \phi_1=\phi_2+\frac {\partial
\phi_3}
{\partial y}
$$

In the general case UToda$(m_1,m_2)$ substitution connect
$2^{m_1+m_2-1}$ initial functions with the same number of the final
ones.

The remarkable property of UToda$(m_1,m_2)$ mappings consist in their
integrability. This means that corresponding to them symmetry
equation
possesses the infinite number of nontrivial solutions. Each solution
of
the symmetry equation initiate the completely integrable system
invariant with
the respect of transformation of UToda$(m_1,m_2)$ substitution
\cite{FL}. By
this property all such systems are united into corresponding
integrable hierarchy.

\section{Evolution parameters and the Integrable Hierarchies}

In this section we introduce the parameters of evolution and
explain the way of constructing the evolution type systems of
equations
together with their explicit soliton-like solutions. All such
constructed
systems are invariant with the respect to UToda$(m_1,m_2)$
substitutions
and belong to the integrable hierarchy with the same title.

We will assume that arbitrary functions, which enter in equations for
$M^{\pm}$ elements (\ref{ME}) in their turn depend on some additional
$t_{n_2},\bar t_{n_1}$ (left-right time) parameters
in such a way that $M^{\pm}$ elements satisfy additional equations
selfconsistent with (\ref{ME}):
\begin{equation}
\frac {\partial M_+}{\partial \bar t_{n_1}}=\sum_{s=0}^{n_1} P^{+s}
M_+,
\quad \frac {\partial M_-}{\partial t_{n_2}}=\sum_{s=0}^{n_2}
R^{-s} M_- \label{MET}
\end{equation}
and where now $P^{+s}(y,\bar t_{n_1}),R^{-s}(x,t_{n_2})$ are the
functions of
their arguments taking values in subspaces with $\pm s$ graded
indexes.
Of course, now in (\ref{ME}) dependence on involved in it values is
the
following $A^{+s}(y,\bar t_{n_1}),B^{-s}(x,t_{n_2})$.

Further content of the present section we divide on four parts. In
two first
ones we demonstrate in what connection are equations
(\ref{ME}) and (\ref{MET}) with the systems of integrable hierarchy.
In
the third one we present the explicit solution of the problem of
self-consistency of (\ref{ME}) and (\ref{MET}). In fourth part we
briefly
describe the construction of multi-soliton type solutions of the
systems
of Darboux--Toda (D--T) hierarchy.

To make material more comprehensive and understandable we begin from
the two-dimen\-si\-onal Darboux--Toda hierarchy ( by the name of
corresponding
integrable mapping) all systems of which are invariant with respect
to
transformation of usual Toda chain (UToda$(1,1)$) for which solution
of
the problem was obtained before \cite{DLY} with the help of the
direct
solution of the corresponding symmetry equation.

\subsection{The integrable systems of D--T hierarchy}

Comparing (\ref{ME}) and (\ref{MET}) we conclude, that with respect
to the
pair of the space coordinates $(x,y)$ the corresponding matrix
elements of
group element $K$ satisfy equations of the usual Toda lattice:
\begin{equation}
\frac {\partial^2 \ln < i >}{\partial x\partial y}=\theta_i
\label{TT}
\end{equation}
With respect to argument pair $(\bar t_k, x)$ ( time and space,
respectively)
we have UToda ($k,1$) chain describing by the system:
\begin{equation}
\frac {\partial \bar p^{(s}_i}{\partial x}=
\theta_{i+s}\bar p^{(s+1}_i-\theta_{i-1}\bar p^{(s+1}_{i-1},\quad
\frac {\partial^2 \ln < i >}{\partial x\partial \bar t_k}=
\theta_i \bar p^{(1}_i,\quad 1\leq s \leq (k-1),\quad \bar p^{(k}_i=1
\label{EX}
\end{equation}
And finally with respect to argument pair $(y,t_k)$ we have
UToda($1,k$)
chain with the corresponding equations:
$$
\frac {\partial p^{(s}_i}{\partial x}=
\theta_{i+s} p^{(s+1}_i-\theta_{i-1} p^{(s+1}_{i-1},\quad
\frac {\partial^2 \ln < i >}{\partial y\partial t_k}=
\theta_i p^{(1}_i,\quad 1\leq s \leq (k-1),\quad p^{(k}_i=1
$$

Resolving the equations of zero curvature with respect to remaining
pairs
$(x,t_{n_2})$ and $(y,\bar t_{n_1})$ will be done in the third
subsection.

Here we would like to emphasize that in all three examples above the
explicit
dependence of the functions $p^{(s}_i,\bar p^{(s}_i$ on matrix
elements of
the single group element $K$ is not the same and determines by the
corresponding formulae (\ref{SUP}) of the previous section.

Instead of the general consideration we consider several simplest
examples
from which the situation in the general case become absolutely clear
( we also
restrict ourselves by the choice of the left time parameter).

\subsubsection{k=2}

Integrating over the argument $x$ the first and second equations
(\ref{EX}) we
obtain consequently:
$$
\frac {\partial \ln < i > }{\partial \bar t_k}=\int^x dx' \theta_i
\bar p^{(1}_i=\int^x dx' \theta_i \int^{x'} dx'' (\theta_{i+1}-
\theta_{i-1})
$$
Introducing the functions $v_i={< i+1 > \over < i >},u_i={< i-1>
\over < i >}$, we obtain the following system of equalities for them
($\theta_i=u_i v_i$):
$$
-\frac {\partial u_ i}{\partial \bar t_2}=\int^x dx' \theta_i
\int^{x'}
dx'' (\theta_{i+1}-\theta_{i-1})-\int^x dx' \theta_{i-1} \int^{x'}
dx'' (\theta_i-\theta_{i-2})
$$
\begin{equation}
{}\label{DS}
\end{equation}
$$
\frac {\partial v_ i}{\partial \bar t_k}=\int^x dx' \theta_{i+1}
\int^{x'}
dx'' (\theta_{i+2}-\theta_i)-\int^x dx' \theta_i \int^{x'}
dx'' (\theta_{i+1}-\theta_{i-1})
$$
The last system is exactly Davey-Stewartson one \cite{DS} rewritten
in terms
of discrete transformations shifts.

Indeed, the UToda($1,1$) integrable mapping (\ref{TT}), rewritten in
the terms 
of $(u,v)$ functions, takes the form:
\begin{equation}
\tilde u= v^{-1}\quad \tilde v= v(uv- (\ln v)_{xy})\label{DT}
\end{equation}
and was called before as Darboux--Toda integrable mapping \cite{DLY}.

Performing all necessary changes of variables in (\ref{DS}) with the
help
of (\ref{DT}), we come finally to ($u_i\to u,v_i\to v$):
\begin{equation}
-\dot u+u_{yy}+2u\int dx(uv)_y=0\quad \dot v+v_{yy}+2v\int dx(uv)_y=0
\label{20}
\end{equation}
This is exactly Davey-Stewartson system in its original form
\cite{DS}.
In one-dimensional limit - usual nonlinear Schrodinger equation.

In \cite{DLY} it was obtained the sequence of the solutions of
symmetry
equation corresponding to Darboux-Toda integrable substitution and
(\ref{DS})
is one them.

\subsubsection{k=3}

Literally repeating the calculations of the the last subsubsection,
we have
consequently:
$$
\frac {\partial \ln < i > }{\partial \bar t_3}=\int^x dx' \theta_i
\bar p^{(1}_i=\int^x dx' \theta_i \int^{x'} dx'' (\theta_{i+2}
\bar p^{(2}_i-\theta_{i-1}\bar p^{(2}_{i-1})
$$
Substituting into the last expression $\bar p^{(2}_i$ in terms of
$\bar
p^{(3}_i$ and keeping in mind that it is necessary to put $\bar
p^{(3}_i=1$
in this case, we finally obtain for derivative
$\frac {\partial \ln < i > }{\partial \bar t_3}$ expression
consisting of
four terms.
Equations (equalities) for $u_i,v_i$ in its turn contain eight terms
with
three repeated integrals and exactly coincide with those from the
paper
\cite{DLY}.

Now the strategy of calculations in the case of arbitrary $k$ is
absolutely
clear and final result is the same as in the cited paper \cite{DLY}.

We bring to the attention of the reader the fact that constructed
systems
are satisfied under arbitrary choice of the index $i$ in them. This
means
that constructed evolutional-like systems all are invariant with
respect
to transformation of Darboux-Toda substitution (\ref{DT}).

\subsection{The integrable systems of $UToda(m_2,m_1)$ hierarchy}

In the general case as a consequence of the condition of
self--consistency
of the equations (\ref{ME}) and (\ref{MET}) ( for definiteness we
keep in mind
the case of "left" time) with the respect the pair of the "space"
coordinates $(y,x)$ usual $UToda(m_2,m_1)$ substitution (\ref{SUP})
arises
and with the respect of the space--time pair  $(\bar t_k, x)$ the
$UToda(k,m_1)$ ones
( it is necessary in (\ref{SUP}) only change $m_2\to k$).

Explicit solution of both these systems (with fixed ends) are
constructed
from the matrix elements of the single group element $K$ by the rules
of
the previous section. From the corresponding formulae reader can see
that 
the functions $< i >, p^{(s}_i$ are the same (by their dependence on
matrix
elements of $K$) than for $\bar p^{(s}_i$ this dependence is
absolutely
different. To search this fact we will denote $\bar p^{(s}_i$ from
the
space--time system by big letter.

We rewrite $UToda(k,m_1)$ system in useful for us notations (
integrated
by space coordinate $x$ first two its equations):
$$
\bar P^{(r}_i=\int^x dx' \sum_{q=1}^{m_2-r} (\Theta^{+(q-1)}_{i+r}
p^{(q}_{i+r}\bar P^{(q+r}_i-\Theta^{-(q-1)}_{i-1} p^{(q}_{i-q}\bar P^
{(q+r}_{i-q})
$$
\begin{equation}
\frac {\partial \ln (\vc{i} K \ve{i})}{\partial \bar t_k}=\int^x dx'
\theta_i
^{-1} \sum_{p=0,q=0}^{p+q\leq Min\{ k-1,m_1-1 \}} \Theta^{-p}_i
\Theta^{+q}_i
p^{(p+q+1}_{i-p}\bar P^{(p+q+1}_{i-p}
\label{SUPT}
\end{equation}
$$
\frac {\partial p^{(r}_i}{\partial \bar t_k}=\sum_{q=1}^{m_1-r}
(\Theta^
{+(q-1)}_{i+r} \bar P^{(q}_{i+r} p^{(q+r}_i-\Theta^{-(q-1)}_{i-1}\bar
P^{(q}
_{i-q} p^{(q+r}_{i-q})
$$
Keeping in mind the condition $\bar P^{(k}_i=1$ and "nilpotent"
character of
the first system of equations we can resolve the last one and obtain
the
explicit expressions for all $\bar P^{(s}_i$ functions in form of
repeated integrals on space coordinate $x$ with integrand functions
always be
some functionals of the functions $< i >, p^{(s}_i$. Substituting
obtained in
such way expressions for $\bar P^{(s}_i$ functions into two last
systems of
(\ref{SUPT}), we find the explicit form of derivatives of $< i >,
p^{(s}_i$
functions with respect to the time argument.

In the same way it is possible to resolve first system of equations
of
$UToda(m_2,m_1)$ substitution with respect to $\bar p^{(s}_i$
functions
as functionals of the same type, as has made above on $< i >,
p^{(s}_i$ 
functions.
Now in the last iteration procedure $\bar p^{(m_2}_i=1 (!)$.
Knowledge
of the time derivatives of $< i >, p^{(s}_i$ functions allows to
reconstruct
the time derivatives of all $\bar p^{(s}_i$ functions.

So we have the time derivatives of all functions involved into
$UToda(m_2,
m_1)$ substitution in terms of functionals of themselves and their
discrete shifts of the correspondingly (limited) order. But with the
help of
equations of $UToda(m_2,m_1)$ mapping (\ref{SUP}) it is always
possible
to present these shifts in term of exactly of $2^{m_1+m_2-1}$ initial
functions and its derivatives up to the definite order.

So we have obtained the system of equalities between the time
derivatives of
$2^{m_1+m_2-1}$ functions expressed in sufficiently cumbersome
functional form
(nonlinear and nonlocal simultaneously) on their space derivatives.

Reminding about the way of obtaining we can consider the last system
as
completely integrable one with known sequence of its soliton--like
solutions.

\subsection{Solution of the Nilpotent chain system}

Up to now we have not resolved only two pairs of nilpotent systems on
the
point of their selfconsistency. For definity let us consider the
time-space pair $(y,\bar t_k)$ and restrict ourselves by the case of
$UToda(1,k)$ substitution. We rewrite the corresponding systems ( as
the
combination of the components of the equations (\ref{ME}) and
(\ref{MET})):
\begin{equation}
(m_+)_{\bar t_k}=((\bar g_0)^{-1}(\bar g_0)_{\bar t_k}
+\sum_{s=1}^kp^{(+s})
m_+\quad (m_+)_y=((\bar g_0)^{-1}(\bar g_0)_y +I^{(+1})m_+ \label{E1}
\end{equation}
where $\bar g_0\equiv \exp \sum_{k=1}^r (h_k \tau_k)$.
Maurer-Cartan identity applied to this pair of equations (\ref{E1}),
is
equivalent to chain like system for unknown functions
$\bar g_0,\quad  \pi^{(+s}=\bar g_0 p^{(+s} \bar g^{-1}_0$
(differentiation
with respect to argument $\bar t_k$ we denote by $\dot {}$, with
respect to
$y$ by ${}'$ and for a time put $\bar g_0\to g$):
\begin{equation}
(\pi^{(+1})'=\dot {g I^{(+1} g^{-1}},\quad
(\pi^{(+s})'=[\pi^{(+(s-1)},
g I^{(+1} g^{-1}] \label{E3}
\end{equation}
where $\pi^{(+k}_i=1$ and $I^{(+s}$ means that in $p^{(+s}$ all
$p^{(+s}_i=1$.

Reminding that $p^{(+s}$ ( and correspondingly $\pi^{(+s}$)
may be presented as a direct sum of components, we rewrite (\ref{E3})
in
component form:
$$
(\pi_i^{(+1})'=\dot {(g_i g_{i+1}^{-1})}\quad (g_ig_{i+k}^{-1})'=
\pi^{(k-1}_i
(g_{i+k-1} g_{i+k}^{-1})-(g_i g_{i+1}^{-1})\pi^{(k-1}_{i+1}
$$
$$
(\pi_i^{(+s})'=\pi^{(s-1}_i (g_{i+s-1} g_{i+s}^{-1})-(g_i
g_{i+1}^{-1})
\pi^{(s-1}_{i+1}\quad 2 \leq s \leq (k-1)
$$
And at the last after identification $G_i\equiv g_i g_{i+1}^{-1}=
\exp (\tau_{i+1}-2\tau_i+\tau_{i-1})$, we come to the
final system of equations for determining of the unknown functions  $
G_i,
\pi^{(+s}_i$:
\begin{equation}
(\pi_i^{(+1})'=\dot {G_i}\quad (G_i G_{i+1}... G_{i+k-1})'=
\pi^{(k-1}_i G_{i+k-1}-G_i \pi^{(k-1}_{i+1}\label{NE}
\end{equation}
$$
(\pi_i^{(+s})'=\pi^{(s-1}_i G_{i+s-1} -G_i \pi^{(s-1}_{i+1}\quad
2 \leq s \leq (k-1)
$$
In spite of very complicate on the first look structure of the last
chain-like system the solution of it it is possible to find in
explicit form.

For this purpose let us consider the following linear equation for
unknown
function $X$:
\begin {equation}
\dot X=X^{(k}+A^{(2} X^{(k-2}+.....+A^{(k} X \label{LE}
\end{equation}
where $A^{(s}$ are arbitrary functions of two arguments $y,\bar t_k$.

The following assertion takes place:

Let
$$
X_1=\phi_1, \quad X_2=\phi_1 \int^y dy' \phi_2(y'),\quad
X_3=\phi_1 \int^y dy' \phi_2(y') \int^{y'} dy'' \phi_3(y''), .....
$$
different solutions of linear equation (\ref{LE}) presented in
Frobenious-
like form. Then the solution of chain-like system
(\ref{NE}) may be written in terms of these solutions as follows:
\begin{equation}
G_i=\phi_i\equiv {Det_{i+1}( X )\over Det_i ( X )},\quad
X_{p,q}=\frac 
{\partial^{p-1} X_q}{\partial y^{p-1}}   \label{ASER}
\end{equation}
where matrix $X$ by the form coincides with the matrix of Vronsky
determinant.
The functions  $\pi_i^{(+s}$ in its turn may be expressed in terms of 
repeated integrals with the known integrands $G_i,\dot G_i$ after
consequent 
integration of corresponding "nilpotent" system (\ref{NE}) for them.

Moreover for only ones necessary for further consideration functions
$g_i$ 
we obtain: 
\begin{equation}
g_i\equiv (\bar g_0)_i=Det^{-1}_i ( X ) \label{FE} 
\end{equation}

To prove this assertion in the general form we will substitute by the
detailed consideration of two examples from which the general case
become
absolutely clear.

\subsubsection{k=2}

The system (\ref{NE}) takes the form:
\begin{equation}
(\pi_i^{(+1})'=\dot {G_i}\quad (G_i G_{i+1})'=
\pi_i^{(+1} G_{i+1}-G_i \pi^{(+1}_{i+1}\label{NEI}
\end{equation}
or the system of chain-like equations for only one unknown function
$G_i$:
\begin{equation}
(G_i G_{i+1})'=\int dy \dot {(G_i)}  G_{i+1}-G_i \int dy \dot
{(G_{i+1})}
\label{NEII}
\end{equation}

Now let us consider the equation (\ref{LE}) for $k=2$. For
$\phi_{1,2}$
we obtain consequently:
$$
\dot {\phi_1}=\phi''_1+A^2 \phi_1,\quad \dot {\phi_2}=(2\phi^{-1}_1
\phi_2
+\phi_2)'
$$
The fact that $X_3=\phi_1 \int^y dy' \phi_2(y')x_2 \int^{y'} dy''
\phi_3(y'')$ is equivalent to the substitution $\phi_2 \to \phi_2
\int dy 
\phi_3$. After trivial algebraical manipulations we come to equality:
$$
\phi_2ú \int dy \dot {\phi_3}-\int dy \dot {\phi_2} ú
\phi_3=(\phi_2\phi_3)'
$$
Further substitution $\phi_3 \to \phi_3 \int dy \phi_4$ ( $X_4$ is
also the
solution of the equation (\ref{LE})!) leads to the result:
$$
\phi_3ú \int dy \dot {\phi_4}-\int dy \dot {\phi_3} ú
\phi_4=(\phi_3\phi_4)'
$$
By induction it is not difficult to show that each two consequent
functions
$\phi$ are connected by relation:
$$
\phi_iú \int dy \dot {\phi_{i+1}}-\int dy \dot {\phi_i} ú \phi_{i+1}
=(\phi_i\phi_{i+1})'
$$
Comparison of the last relation with (\ref{NEII}) leads to
conclusion, that 
in the case under consideration:
$$
G_i=\phi_i,\quad \pi^{(+1}_i=\int dy \dot {(G_i)}
$$
and the assertion (\ref{ASER}) is proved.

\subsubsection{k=3}

In this case system (\ref{NE}) takes the form:
\begin{equation}
(\pi_i^{(+1})'=\dot {G_i}\quad (G_i G_{i+1}G_{i+2})'=
\pi_i^{(+2} G_{i+2}-G_i \pi^{(+2}_{i+1}\label{NEIII}
\end{equation}
$$
(\pi_i^{(+2})'=\pi^{(+1}_i G_{i+1} -G_i \pi^{(+1}_{i+1}\quad
$$
or after excluding all functions $\phi^{(1,2}_i$ we come to a
chain ( nonlinear and nonlocal simulteneously) system of equations
for
functions $G_i$:
$$
G_{s+2}(y)[\int^y dy'G_{s+1}(y') \int^{y'} dy'' \dot G_s(y'')-
\int^y dy'G_s(y') \int^{y'} dy'' \dot G_{s+1}(y'')]-
$$
\begin{equation}
{}\label{NE3}
\end{equation}
$$
G_s(y)[\int^y dy'G_{s+2}(y') \int^{y'} dy'' \dot G_s(y'')-\int^y
dy'G_{s+1}
(y')\int^{y'} dy'' \dot G_{s+2}(y'')]=(G_sG_{s+1}G_{s+2})'
$$

Now let us consider equation (\ref{LE}) in the case $k=3$:
\begin {equation}
\dot X=X'''+A^2 X'+A^3 X \label{LE3}
\end{equation}
Substituting in it the form of the solution $X_1, X_2$ proposed by
the 
assertion (\ref{ASER}) we obtain:
$$
\dot \phi_1=\phi'''_1+A^2 \phi'_1+A^1 \phi_1,\quad \dot \phi_2=
(3\phi^{-1}_1\phi''_1\phi_2
+3\phi^{-1}_1\phi'_1\phi'_2+\phi''_2+A^2\phi_2)'
$$
The fact that $X_3$ is also solution of the same equation equivalent
to the change $\phi_2\to \phi_2\int \phi_3$ and leads to the
following
equality:
$$
\phi_2 \int \dot \phi_3-\phi_3 \int \dot \phi_2=
(3\phi^{-1}_1\phi'_1\phi_2\phi_3 +2\phi'_2\phi_3+\phi_2\phi'_3)'
$$
And at last after substitution $\phi_3\to \phi_3\int \phi_4$, which
is
equivalent to proposition that $X_4$ is also solution of the same
linear
equation, we come to equality of our interest:
$$
\phi_2(y)[\int^y dy'\phi_3(y') \int^{y'} dy'' \dot \phi_4(y'')-
\int^y dy'\phi_4(y') \int^{y'} dy'' \dot \phi_3(y'') ]-
$$
$$
[\int^y dy'\phi_2(y') \int^{y'} dy'' \dot \phi_3(y'')-\int^y
dy'\phi_3(y')
\int^{y'} dy'' \dot
\phi_2(y'')]\phi_4(y)=(\phi_2(y)\phi_3(y)\phi_4(y))'
$$
By the induction this equality can be continued on all three
arbitrary
consequent functions $\phi_s(y),\phi_{s+1}(y),\phi_{s+2}(y)$, solving
in
explicit form the system (\ref{NE3}) and proving the proposed above
assertion 
(\ref{ASER}).

\subsection{Multi-soliton like solutions of D--T hierarchy}

From the results of the last section it follows the deep connection
between
the general solution of two-dimensional Toda lattice with fixed ends
(what is equivalent to exploiting of finite-dimensional $A_n$
algebra) and
the particular soliton-like solutions of the systems of D--T
hierarchy.

To obtain such kind of solutions of the systems of D--T hierarchy,
which
may be enumerated by index $k$ in (\ref{LE}),it is necessary only
substitute
into the general solution of Toda chain encoded in functions $< i >$
instead 
of arbitrary functions $\bar g_0(y),(g_0(x))$ the functions $\bar
g_0(y,
\bar t_k)$ from (\ref{FE}) and corresponding expression for
$g_0(x,t_k)$.

As a corollary each pair of functions
$$
u^k\equiv u^k_i={< i-1 >_k\over < i >_k},\quad v^k\equiv v^k_i={< i+1
>_k
\over < i >_k}
$$
satisfy the k-th system of the integrable D--T hierarchy
simultaneously
with respect to left and right time parameters. The first nontrivial
example 
of this hierarchy $(k=2)$ is the D--S system (\ref{DS}) in the form
of the 
discrete shifts or (\ref{20}) in the form of the usual derivatives.

Some additional consideration it is necessary to extract the
solutions 
invariant with respect to some inner authomorphism of the problem or
more 
precisely of D--T substitution ( in this connection see \cite{LG}),
\cite{LY},\cite{GR}).

\section{Possible generalization on the multi-dimensional case}

Let for some gradings ( this can be satisfied far not always) it is
possible represent "lagrangians" $L^{(+m_1}(y),L^{(-m_2}(x)$ from
(\ref{ME})
in the block form $L^{(+m_1}=\sum_s L^{(+m_1}_s$, where elements of
different
blocks are mutually commutative $
[L^{(+m_1}_s(y),L^{(+m_1}_{s'}(y')]=0$.
And the same may be true in some decomposition is true with respect
to the 
second lagrangian.

Then the solution of (\ref{ME}) may be presented in the form of the
product
of $p_1$ ($p_1$ is the number of blocks of the above type) mutually 
commutative factors:
\begin{equation}
M^+(y)=\prod_s^{p_1} M^+_s(y) \label{M}
\end{equation}

Let us instead of $M^+(y)$ consider the new element
$M^+(y_1,...y_{p_1})$
depending on  $p_1$ arguments $y_s$, determined by relation:
$$
M^+(y_1,...y_{p_1})=\prod_s^{p_1} M^+_s(y_s)
$$
The same procedure it is possible to realize with $M^-$ and change it
on the
group-valued function on $p_2$ independent arguments:
$$
M^-(x_1,...x_{p_2})=\prod_s^{p_2} M^-_s(x_s)
$$
Now we construct as in the second section the group-valued element
$$
G=N_-M_+=g_0 N_+M_-
$$
and with respect to $p_1$ and $p_2$ algebra-valued functions
$$
G_{x_i}G^{-1}=(N_-)_{x_i}N_-^{-1}=\sum_s^{m_2} R^{(-s}_i(x,y),\quad
1\leq s\leq m_2
$$
\begin{equation}
{}\label{LAA}
\end{equation}
$$
G_{y_m}G^{-1}=(g_0)_{y_m}g_0^{-1}+g_0(N_+)_{y_m} N_+^{-1}(g_0)^{-1}=
(g_0)_{y_m} g_0^{-1}+\sum_s^{m_1} R^{(+s}_m(x,y),\quad 1\leq s\leq
m_1
$$
come to the same conclusions as it was done before with respect to
the case 
of the single x and y arguments.

Since to go further it is necessary to understand from pure
algebraical
point of view the possible block structure of described above type as
a direct
corollary of the properties of the algebra and chosen grading in it.
Now we are not ready to solve this problem in the whole measure.

\section{Outlook and further perspectives}

The main result of the present paper consists in proposition that 
the theory of integrable systems in $(1+2)$ dimensions (and of course
in 
$(1+1)$ case as a direct reduction of the previous one) is nothing
more than 
the equations of equivalence from the representation theory of
semisimple 
algebras and groups encoded in some nontrivial way.  It is true at
least 
in the framework of integrable substitutions considered here.

We have the chain of the following consequent steps: 
Graded Lee algebras -- Representation theory --
Integrable mappings (substitutions) and, at last, evolution--type
Hierarchies
of Integrable systems together with their (soliton-like) solutions. 
This chain after its consequent realization is nothing more than 
the theory of integrable systems belonging to the same hierarchy 
in $(2+2)$ dimensions (keeping in mind 
the existence of the "left" and "right" time parameters). 

On the other side, the theory of integrable systems, in particular,
in $(1+1)$ and $(1+2)$ dimensions, was up to the latest time,
the independent branch of mathematical physics with its own technique
and methods of investigations \cite{ZFOD}. This means only that there
are
many independent methods for investigation of the representation
theory of the
semisimple algebras and groups. The methods applied in the theory
of integrable  systems were ones among many other possible ones.

The researches worked in this area have rediscovered for many times
and
by absolutely independent methods the different forms of the two
Jacobi identities (\ref{recrel}) and (\ref{J2}) and numerous
corollaries
from them. Of course, in the case of $A_n$ algebra and the principal
grading
in it it is possible to  perform all of this job
in the language of Jacobi identities in the determinant form, which
on the first look have no connection to representation theory and may
be
performed without any mention about its existence.

In context of the material of the present paper it arised many other
problems some ones of which we want to emphasize.

First of all ( and this is sufficiently obvious) to try to generalize
the
presented construction to the case of arbitrary semisimple algebra
together
with the principle (embedding) grading in it. This problem may be
identified as the case of Abelian Integrable Mapping in the framework
of
semisimple algebras.  For solution of this problem it is necessary
the more
detailed information about the structure of the algebra in the case
of
the existence of the repeated roots as for its fundamental
representations
in the case of arbitrary semisimple algebra. In this connection see
Appendix.

Secondly the problem of generalization to the case of arbitrary
grading
of semisimple algebras arises. In general case the subspace with zero
graded
index become noncommutative algebra by itself and it leads to
additional
technical difficulties. The typical example is the simplest case of
so
called matrix Toda lattice considered from the different points of
view in
the papers of the different groups of authors
\cite{RS},\cite{LY},\cite{GR}.

And the last problem in framework of our comments, how  this
construction works in the case of Lie algebras of the general
position and
to what kind of integrable substitutions it leads?
This is a subject of additional and non trivial ( as it possible to
assume)
investigation.

\noindent{\bf Acknowledgements.}

Author is indebted to the Instituto de Investigaciones en
Matem'aticas
Aplicadas y en Sistemas, UNAM and especially  to  its  director
Dr. I.  Herrera  for beautiful conditions for his work.
Author freundly thanks N. Atakishiyev, S.M. Chumakov and
K.B. Wolf for permanent discussions in the process of working on this
paper.
This work was done under partial support of Russian Foundation of
Fundamental
Researches (RFFI).

\section*{Appendix: GToda($ 2,2;s,\bar s$) lattice}

\setcounter{equation}{0}
\def\theequation{A.\arabic{equation}}
\vspace{0.5cm}

The tittle of this section is decoded as follows: we consider
arbitrary
semisimple algebra together with the principle grading in it and in
the main
equations (\ref{ME}) restrict ourselves by the choice $m_1=m_2=2$.
The sense
of additional parameters become clear from what follows (see
\cite{dls}).

In this case "Lagrangians" take the form
$$
L^2_{\pm}=(h\tau_{\pm})+\sum X^{\pm}_{\alpha} \phi^{\pm
1}_{\alpha}+{1\over 2}
\sum [X^{\pm}_{\alpha}, X^{\pm}_{\beta}] \phi^{\pm 2}_{\alpha,\beta}
\quad
\phi^{\pm 2}_{\alpha,\beta}=-\phi^{\pm 2}_{\beta,\alpha}
$$
The action of $L^2_-$ on the state vector $\ve{i}$ is as follows:
$$
L^2_- \ve{i}=(\phi^{-1}_i+\sum_{k_{\alpha,i}\neq 0}
\phi^{-2}_{\alpha,i}
X^-_{\alpha})X^-_i \ve{i}
$$
Calculation of derivatives of $\alpha^1,\bar \alpha^1$ (\ref{A}) by
the same
technique as in the main text ( section 4) leads to the result:
$$
\frac {\partial \bar \alpha^1_i}{\partial x}=< i >^{-2}
(\phi^{-1}_i+\sum_{k_{\alpha,i}\neq 0} \phi^{-2}_{\alpha,i}
(X^-_{\alpha})_
{left}) {\prod}^r_{j=1,i\neq_j} \vc{j} G \ve{j}^{-K_{ij}}=
$$
$$
\theta_i ((\phi^{-1}_i-\sum_j k_{j,i} \phi^{-2}_{j,i}
\alpha^1_j),\quad
\theta_i={\prod}_j \vc{j} G \ve{j}^{-K_{ij}},\quad K_{j,j}=2
$$
After introductions the functions:
$$
p^1_i=\theta_i^{-1}\frac {\partial \bar \alpha^1_i}{\partial x}\quad
\bar p^1_i=\theta_i^{-1}\frac {\partial \alpha^1_i}{\partial y}
$$
we obtain the following closed system of equalities for these
functions
together with $< i >$:
$$
\frac {\partial p^1_i}{\partial y}=\sum_j K_{j,i} \phi^{-2}_{j,i}
\theta_j
\bar p^1_j\quad
\frac {\partial \bar p^1_i}{\partial x}=\sum_j K_{j,i}
\phi^{+2}_{j,i}
\theta_j p^1_j
$$
\begin{equation}
{}\label{MFE}
\end{equation}
$$
\frac {\partial^2 \ln < i >}{\partial x\partial y}=\theta_i p^1_i
\bar p^1_i-
\theta_i \sum_j K_{j,i} \theta_j\phi^{-2}_{j,i} \phi^{+2}_{j,i}
$$
The second mixed derivatives of $\ln < i >$ is calculated without any
difficulties by the same way as first equations of (\ref{MFE}).

Comparing (\ref{SUP}) ( in the case $m_1=m_2=2$) with (\ref{MFE})
after
substitution in the last system the Cartan matrix of $A_n$ algebra,
shows the
whole identity of this systems under additional choice of arbitrary
functions
$\phi^{\pm}_{j,i}=1$.

For simplicity in the main text of the paper we put $\bar p^1_i=\bar
p^1_i=
1$. In fact this is some additional assumption and we want now to get
rid of
it.

Indeed the main equations (\ref{ME}) are obviously invariant with
respect to
gauge transformation with the group element $\exp (h\tau_{\pm})$
($\tau_+\equiv \tau_+(y)\quad \tau_-\equiv\tau_-(x)$). With the help
of such
transformations all $\phi^{\pm 2}_{\alpha,\beta}$ may be evaluated to
a
constant values. Let us work in such gauge, where they take zero and
unity
values in arbitrary order and denote such sequences of these
parameters by
the symbol $\bar s, s$. The last finally explain the notation GToda($
2,2;s,
\bar s$) in the title of this Appendix. The arising systems are
essentially
different as by the form of the equations by itself also as by the
form of
their general solutions (see in this connection \cite{dls}).

\end{document}